\documentclass[aps,pre,twocolumn,groupedaddress]{revtex4}

\usepackage{graphicx}
\usepackage{dcolumn}
\usepackage{bm}
\usepackage{longtable}
\usepackage{xcolor,soul}
\sethlcolor{yellow}

\begin{document}


\title{A Geometrically Exact Treatment of Percolation Through Voids around Faceted
Regular and Structurally Disordered Grains} 


\author{D. J. Priour, Jr}
\affiliation{Department of Physics \& Astronomy, Youngstown State University, Youngstown, OH 44555, USA}
\altaffiliation{}


\date{\today}

\begin{abstract}
Fluid and charge flow through interstitial volumes among impermeable randomly placed 
grains in porous materials ceases to occur at a critical concentration where networks of 
void volumes are disrupted at macroscopic scales.  This critical density for void 
percolation can be difficult to calculate due to the irregular shape of the 
void regions.  We develop and implement a geometrically exact method, scaling only 
linearly in the system volume, for identifying the shape and size of contiguous 
voids.  In this manner, we calculate percolation thresholds for both grain cluster 
percolation (where system spanning networks of overlapping grains begin
to appear with increasing density) and void percolation at much higher grain 
concentrations where networks of interstitial volumes no longer exist on 
macroscopic scales.  For both the former and the latter, we calculate 
critical concentrations for inclusions in the shape of the Platonic solids (as well as 
truncated icosahedra) for both aligned and randomly oriented grains. 
In the case of critical densities for void percolation, the accuracy of our 
results is significantly improved relative to prior benchmarks.
We also incorporate 
structural disorder of inclusions by considering impermeable grains in the form of 
cubes subject to a series of randomly placed and oriented fracture planes to mimic
aggressively fractured inclusions found in nature.  As the 
number of sustained slices becomes large, we find that the critical porosity 
for void percolation tends to 5\%
\end{abstract}

\maketitle
\section{Introduction}
Though well characterized in discrete lattices~\cite{stauffer}, percolation transitions in systems 
based on continuum geometries such as randomly placed grains impermeable to fluid flow or 
charge transport lack a well defined lattice.  Nevertheless, in the thermodynamic limit, 
whether the system percolates (is impermeable to fluid or charge) is determined by the 
concentration $\rho$ of barrier particles per unit volume.  In this work, 
we use the dimensionless quantity $\eta = \rho v_{\mathrm{B}}$, with $v_{\mathrm{B}}$ being the
volume of the impermeable inclusions.  In cases in which grains are polydispersed, as in structurally 
disordered inclusions, we instead operate in term of $\eta = \rho \langle v_{\mathrm{B}} \rangle$ 
where $\langle v_{\mathrm{B}} \rangle$ is the disorder averaged grain volume.  

Theoretical studies of percolation transitions in continuum geometries may be subdivided into 
those seeking the threshold density for the appearance of system-spanning clusters of 
overlapping inclusions or calculations of the much higher concentration threshold for the elimination of interstitial void 
volumes navigable on macroscopic scales.  We henceforth refer to the former as the grain cluster transition
and the latter as the void percolation transition with $\eta_{c}^{\mathrm{GC}}$ and $\eta_{c}^{\mathrm{V}}$ 
used both to indicate critical concentrations and to serve as labels for the transitions themselves for grain cluster and void 
percolation respectively.  

In general, a geometrically exact description of the void volumes themselves has proven elusive, though 
discretization schemes have been applied in some cases~\cite{martys,maier,Yi1,Yi2,koza}, while Voronoi networks have been brought to 
bear in the case of systems made up of randomly placed spheres~\cite{elam,marck,rintoul,klatt}.  Stochastically driven simulations involving 
virtual tracer particles infiltrating interstitial volumes~\cite{bauer,beijeren,hofling,spanner,hofling2,kammerer,spanner2,djpriour,djpriour2,ballow}
 have been used to calculate critical parameters 
such as $\eta_{c}$, but have not elucidated the irregularly shaped voids themselves in a deterministic fashion.
A recent work considers what in our case are $\eta_{c}^{\mathrm{GC}}$ and $\eta_{c}^{\mathrm{V}}$ as endpoints of the bicontinuous 
regime of various two-phase continuum systems with an analytical framework for estimating critical densities for the 
onset and termination of the bicontinuous phase~\cite{Skolnick}.

In this work, we develop a technique to directly identify interstitial volumes in a geometrically exact manner
with a computational cost which scales only linearly in the system volume.  
We calculate both $\eta_{c}^{\mathrm{GC}}$ and $\eta_{c}^{\mathrm{V}}$ 
for all grain geometries that we consider, but the impetus for this effort and a primary focus herein is the 
aim of improving the accuracy of $\eta_{c}^{\mathrm{V}}$ relative to prior benchmarks in the literature.  In fact, 
we reduce Monte Carlo statistical errors in the 
case of void percolation results from on the order of 0.3\% to one part in $10^{3}$ or less.  Formerly,
differences in $\eta_{c}^{\mathrm{V}}$ for aligned versus randomly oriented Platonic solids that this study reveals 
were obscured by Monte Carlo statistical errors for all cases apart from 
cube-shaped grains.  However, in this work, improvements in the precision of $\eta_{c}^{\mathrm{V}}$ 
owing to the deterministic approach to identifying void volumes allow us to resolve as 
distinct the critical densities for all but the case of the quasi-spherical semi-regular truncated icosahedra. 

With an intent to consider porous media with a closer resemblance to what one encounters in the geological context, we 
consider cases in which the constituent grains are not uniform shapes, a condition we term structural disorder.
As an example of structurally disordered grains, we consider the case of irregular fragments formed from cubes 
subject to a succession of randomly placed and oriented fracturing planes.  We consider a wide range of mean numbers 
of sustained slices per cube, which we use as a metric of the strength of the structural disorder.  
We find that with many accumulated slices, the critical porosity fraction (i.e. $e^{-\eta_{c}^{\mathrm{V}}}$)
saturates at 5\%.

\section{Methods}

In this work, we describe and present results for a method which efficiently (i.e. with computational costs  
scaling only linearly in the system volume) and directly identifies void regions.  We achieve this by 
explicitly finding the boundary surfaces among interstitial volumes and impermeable inclusions. In this manner, 
as we discuss in more detail in this section, one may calculate both $\eta_{c}^{\mathrm{GC}}$ and $\eta_{c}^{\mathrm{V}}$.
The former occurs at low grain concentrations where overlapping grains clusters begin to span the system, 
and where the boundary surfaces, which ensheathe the grain clusters,  begin to percolate as well.  On the other
hand, the void percolation transition occurs at much higher grain concentrations where with increasing density interstitial volumes 
are disrupted to the degree that they are no longer navigable on macroscopic scales. The boundary surfaces, tunnels which now line
voids, also cease to percolate at $\eta_{c}^{\mathrm{V}}$ with increasing $\eta$.

As noted earlier, we use the dimensionless parameter $\eta = \rho \langle v_{\mathrm{B}} \rangle$ with the polyhedral 
inclusion volume obtained from~\cite{polyvolume} for the monodispersed platonic solids and the semi-regular truncated
icosahedron. In the case of structural disorder, one obtains the mean grain volume $\langle v_{\mathrm{B}} \rangle$ 
a posteriori by averaging over the volumes of many structurally disordered inclusions.

To minimize finite size effects and to circumvent artifacts in which spurious large boundary surfaces form for 
$\eta > \eta_{c}^{\mathrm{GC}}$ in the case of free boundary conditions, we use periodic boundary conditions.  
For the sake of computational efficiency, we partition the simulation volume into small cube-shaped voxels 
(similar to Verlet cells~\cite{bruin}).  Each of these small cells, with edges of unit length, is populated with
randomly placed impermeable grains where the number of inclusions in each voxel is determined by sampling
Poissonian statistics. The constituent grains placed in this manner are circumscribed by spheres of unit radius,
though they need not be tangent to the sphere (e.g. the tangency condition generally does not hold in the case of
structurally disordered grains due to the truncation of material by the slicing planes).

In this work, we consider both the cases of aligned and randomly oriented polyhedra.  In the case of randomly
oriented inclusions, constituent planes of the faceted solid are defined in terms of three local axes $\hat{u}_{1}^{'}$,
$\hat{u}_{2}^{'}$, and $\hat{u}_{3}^{'}$.  As is described elsewhere~\cite{djpriour2}, $\hat{u}_{1}^{'}$ and $\hat{u}_{2}^{'}$ are chosen
stochastically, while the third axis is given by $\hat{u}_{3}^{'} = \hat{u}_{1}^{'} \times \hat{u}_{2}^{'}$.

A key element to providing a geometrically
exact description of void volumes is locating each of the vertices on the surface of the interstitial regions.  We find
these vertices in an efficient manner by focusing on the planar faces of the constituent impermeable
inclusions. In general, vertices are defined by the intersection of three planes.  However, in highly symmetric cases,
circumstances arise in which more than three planes coincide in a single vertex. The two examples for which this occurs in
this work are octahedra and icosahedra, where polyhedron vertices mark the intersection of four and five facets, respectively.
To restore the condition of three planes meeting in a vertex, we introduce additional planes to truncate the vertices of the
octahedra and the icosahedra.  The new planes are positioned a distance $d_{0} - \epsilon$ from the polyhedron
center. Here, $d_{0}$ is the (unit distance) to the orginal vertex; with
$\epsilon = 10^{-7}$, only a small portion of the parent polyhedron is truncated (on the order of $10^{-21}$
of the original volume), and all of our results are converged with respect to $\epsilon$.

We locate the vertices of the surfaces bounding void volumes in three stages.
We first characterize each face of the faceted inclusion by finding its edges and vertices.  This may be done with
a priori knowledge of the grain geometry, or by combinatorially sampling sets of three planes making up the
polyhedron.
Here, we
distinguish among the void boundary faces, which may be non-convex and need not be topologically simple and the
(invariably convex) faces of the polyhedral inclusions.
We also eliminate from further consideration any faces entirely engulfed by neighboring grains, a circumstance which
would arise if each of the face's vertices were interior to the inclusion under consideration.
Next, by sweeping over neighboring voxels, one finds each face's set of neighboring faces; a pair of polyhedron faces are deemed
to be each other's neighbors if any of their edges pass through each other's interior, or if they share an edge by
virtue of being adjacent facets in the same polyhedron.

Finally, one considers all sets of three planes which could form a candidate vertex on the surface of the void volume.
To operate efficiently, for each face in the system one sweeps over secondary and tertiary faces in the list of neighbors
of the face under consideration with combinatorial counting used to prevent redundant consideration of three planes.  In brief,
one sweeps through the secondary planes.  For each secondary plane, one then sweeps over the
tertiary planes, after which the secondary plane becomes inactive in the sense that it cannot be considered as a tertiary plane.
After all secondary and tertiary partners have been exhausted for the primary plane, it likewise becomes inactive and cannot
subsequently be selected as either a secondary or tertiary plane.
 In this way, a unique set of distinct neighboring planes is examined once and only once.

Each candidate vertex must be further vetted to determine if it is on the boundary of the void volume, and for this purpose we
impose the constraint that the point under consideration lies on each of the facets associated
with the planes joining to form the vertex.  In addition, the point must not be interior to
any of the grains housing the three facets whose intersection defines the vertex
or to any inclusions with facets neighboring the three facets which form the candidate point.
Vertices which meet these conditions are deemed to lie on the void volume boundary.

Having identified all of the vertices which make up the void volume boundary, we next establish their connectivity to
each other.  In the 3D context, from each point emanate three edges  terminating in neighboring vertices, which are needed
to work out the (often non-convex and topologically non-simple) polygonal faces of the interstitial volume.
With each edge being defined by the intersection of two of the three planes forming the vertex, the neighboring point
we seek is the closest vertex on the edge (i.e. belonging to both edge planes) such that the edge extending to
the point does not cross an empty chasm or pass through a portion of an impermeable grain.

Finally, with neighboring vertices identified for each point on the boundary surface, one enumerates all connected polygons
inscribed in the inclusion faces by randomly choosing a point on a grain face and moving from one
neighbor vertex to the next, such that each new polygon vertex belongs to the inclusion facet under consideration. This process is
iterated until the ring built up in this manner closes on itself. The same sequence of steps is
repeated for another unassigned point on the same inclusion face,
if any exist, to find the next polygon.  One continues in this way until all 
points on the polyhedron facet are exhausted.

One may now be tempted to identify the polygons obtained in this manner with
the actual faces making up the void volume boundary.  However, this is only a valid assumption if the boundary face is
topologically simple, a condition not met (e.g.) if the corner of a grain penetrates the face of another inclusion without
contacting any edges of that face; we therefore need to keep track of the structure of the boundary surface faces, which in some
cases may be of topological genus greater than zero and hence may subsume more than one of the polygons inscribed in the
inclusion face.  For a given polygon, we do this by finding the nearest circumscribing polygon, if any exist, by choosing a
point at random on a segment of a polygon under consideration and constructing a ray along the plane and following it to the
edge of the inclusion face.  In finding the closest ring, if any, which surrounds the polygon under consideration, one
distinguishes among circumscribing and non-circumscribing rings by appealing to the number of times the ray
croses the polygon in question; odd in the case of the former and even for the latter.

The interior polygon and the nearest exterior candidate, if any are present, are part of the same boundary face if in traversing the gap
between the two rings the ray does not pass through an impermeable grain or traverse a chasm of empty space.  One wishes to
prevent the splitting of the vertices belonging to a single void volume boundary into two or more sets of points that would then
erroneously be identified as vertices associated with distinct non-overlapping volumes, which may be achieved by choosing points at random
in both interior and immediate exterior polygons and making them each other's neighbors. Ultimately, we wish to interrogate the entire network of
vertices, and the latter step ensures that each of the vertices associated with the same boundary surface are considered

We find the additional step of
associating polygons for topologically complex boundary faces makes a significant difference only in calculating grain cluster
percolation thresholds, albeit
with the exception of aligned cube shaped grains where the surface facets are invariably topologically simple.
We validate $\eta_{c}^{\mathrm{GC}}$ results obtained in this way by calculating the grain cluster percolation threshold by using 
neighboring faces to work out which grains overlap.  All critical indices, including $\eta_{c}^{\mathrm{GC}}$ are in accord among these
methodologically distinct sets of calculations, and we show results for both efforts.  On the other hand, we also find that the void percolation results 
are not affected to even the slightest degree by the association of polygons with their nearest circumscribing rings if any are present.

\begin{figure}
\includegraphics[width=.4\textwidth]{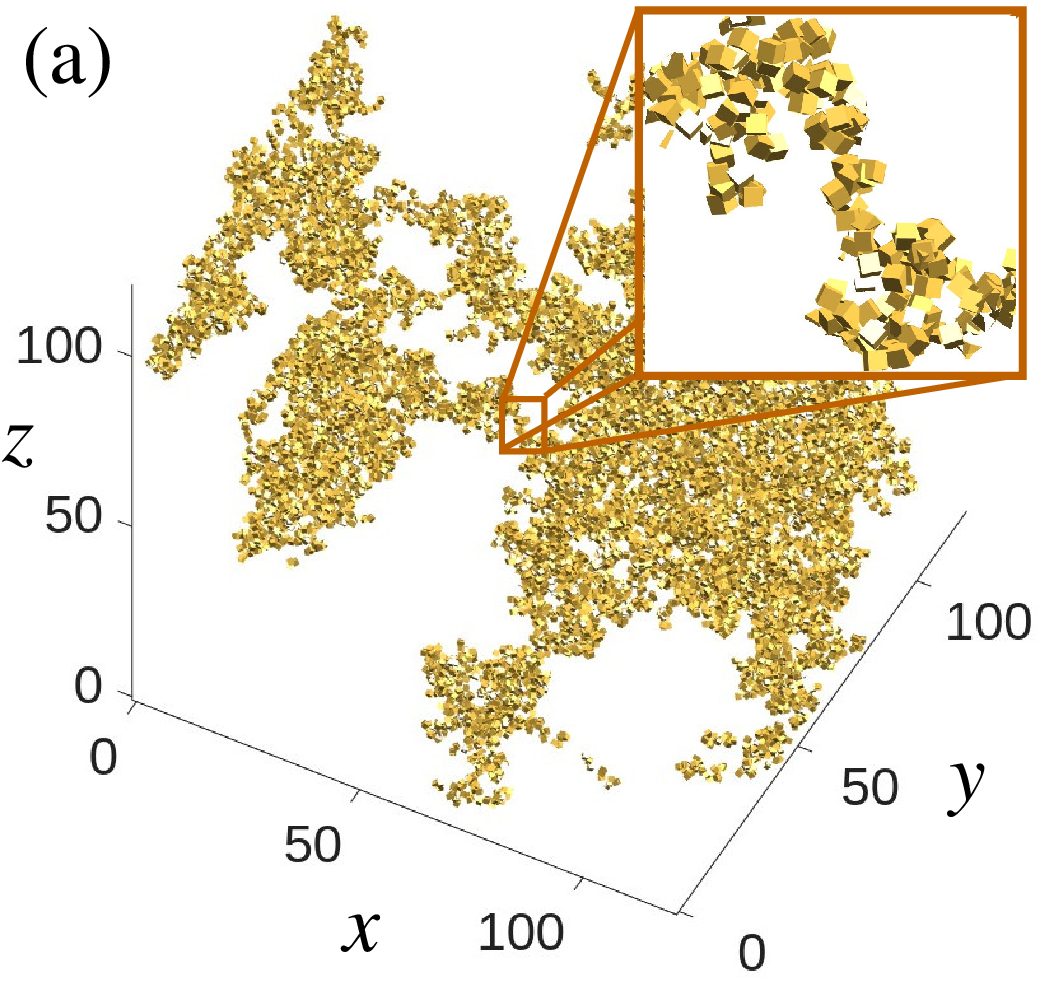}
\includegraphics[width=.4\textwidth]{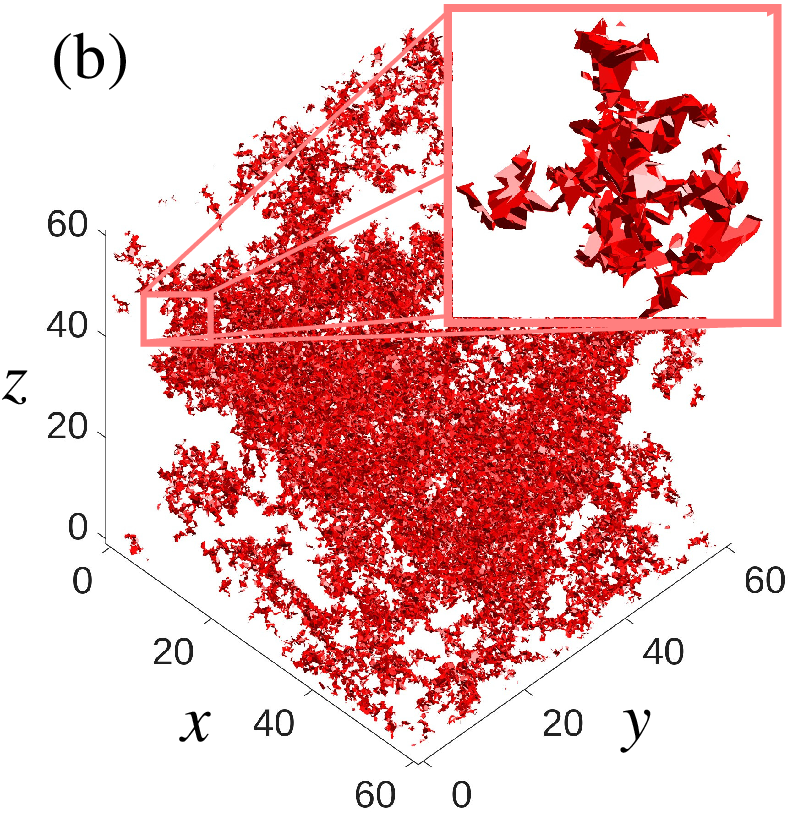}
\caption{\label{fig:Fig1} (Color online) Percolating free surfaces in the
case of randomly oriented cube-shaped grains shown near $\eta_{c}^{\mathrm{GC}}$ in panel (a) and near
$\eta_{c}^{\mathrm{V}}$ in panel (b).}
\end{figure}

In this work, we consider the effective system size to be $L_{\mathrm{eff}} \equiv \langle N \rangle^{1/3}$ where $\langle N \rangle$ is the 
mean number of inclusions present or $L_{\mathrm{eff}} = L \rho^{1/3}$ (with $L$ being the linear dimension of the simulation volume
in voxels).  Our technique can identify all void volume boundary surfaces in a $L_{\mathrm{eff}} = 40$ system
for a particular density $\rho$ in a minute or less on with a single thread on a contemporary CPU. All calculations 
reported on in this work involve averaging over at least $2 \times 10^{4}$ realizations of disorder.  In the case of the 
Platonic solids as well as truncated icosahedra, where our accuracy standard is one part in $10^{3}$ or better, 
we consider system sizes up to at least $L_{\mathrm{eff}} = 40$.   In the case of structurally disordered grains, we 
consider system sizes up to at least $L_{\mathrm{eff}} = 20$ with accuracies of $\eta_{c}$ results on the order of 
a few parts in 1000.
Boundary surfaces ensheathing grain clusters at $\eta_{c}^{\mathrm{GC}}$ are shown in Fig.~\ref{fig:Fig1} 
in panel (a), while void volumes at $\eta_{c}^{\mathrm{V}}$ are shown in panel (b) of Fig.~\ref{fig:Fig1}; 
both instances are for systems made up of randomly oriented cubes. 

All critical indices calculated in this work are based on the disorder averaged percolation fraction $\langle f \rangle$.
The percolation (or lack thereof) of surfaces bounding the void regions marks both 
$\eta_{c}^{\mathrm{GC}}$ and $\eta_{c}^{\mathrm{V}}$.  With increasing inclusion density, there are two 
percolation events involving these boundary surfaces. First, at the grain cluster percolation transition $\eta_{c}^{\mathrm{GC}}$,
boundary surfaces likewise percolate and ensheathe the worm-like clusters that begin span the system.  With further increases 
in $\rho$, interstitial volumes decrease in size and eventually cease to span the system, and the tunnel-like surfaces lining these
voids also no longer percolate.

To determine if a boundary surface percolates in a given case, we use the Hoshen Kopelman algorithm~\cite{hoshen} to identify all 
vertices belonging to the same cluster. We consider a surface to be system spanning if it either wraps around (extending the 
length of the simulation volume and connecting with itself) or extends across the length of the simulation volume without 
joining with itself.  We exploit the inherent subjective nature of the latter to reduce corrections to leading order finite size 
scaling by introducing a tunable parameter $\chi$ such that the condition for percolation in, e.g., the $x$ direction
is $(x_{\mathrm{max}} - x_{\mathrm{min}}) +\chi \geq L$ where $x_{\mathrm{min}}$ and $x_{\mathrm{max}}$ are the locations of 
vertices with the least and greatest $x$ coordinates, respectively.  We impose this condition simultaneously in the 
$x$, $y$, and $z$ directions.  We optimize $\chi$ by insisting that
$\langle f \rangle$ intersections for two distinct pairs of moderately 
sized systems occur at the same grain density. As an example, one may
consider  $\left \{ 8, 12 \right \}$ and $\left \{ 12, 18 \right \}$ for the first and second pairs where 
$L_{1} = 8$, $L_{2} = 12$, and $L_{3} = 18$ with each member of the trio $\left \{ L_{1}, L_{2}, L_{3} \right \}$ being 
50\% greater in size than the last.  
In this context, ``moderately sized'' means that $L_{\mathrm{eff}}$ for $L_{1}$ is at least on the order of 8.

We obtain $\eta_{c}^{\mathrm{GC}}$ and $\eta_{c}^{\mathrm{V}}$ by comparing $\langle f \rangle$ results for
$\left \{ L, \frac{3}{2} L \right \}$ pairs.  In the case of the Platonic solids and truncated icosahedra, we consider 
a minimum of three such pairs with the mean $L_{\mathrm{eff}}$ spanning at least a factor of three among these sets of 
system size pairs.  We endeavor to calculate $\eta_{c}$, $R_{c}$ (the percolation probability $\langle f \rangle$ 
at the critical concentration), and the critical exponent $\nu$ 
associated with the correlation length.  In each case, $\nu$ (up to Monte Carlo statistical error) is 
in accord with $\nu = 0.8764(12)$~\cite{wang} for the 3D percolation universality class  while our $R_{c}$ results in the case of 
$\eta_{c}^{\mathrm{GC}}$ and $\eta_{c}^{\mathrm{V}}$ are compatible with a universal value of 0.83(1).  

\begin{figure}
\includegraphics[width=.45\textwidth]{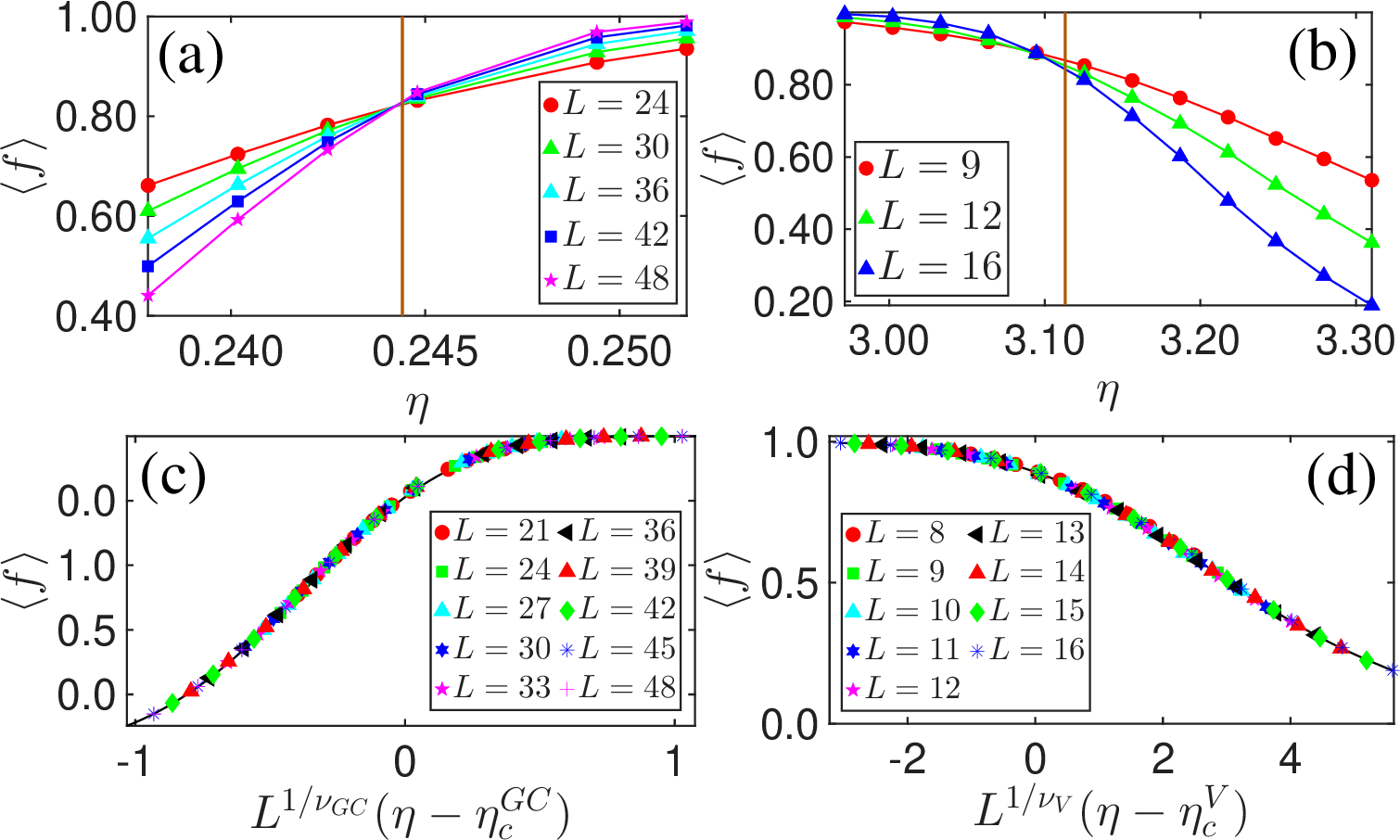}
\caption{\label{fig:Fig2} (Color online) $\langle f \rangle$ for $\eta_{c}^{\mathrm{GC}}$ and $\eta_{c}^{\mathrm{V}}$ are shown in
panel (a) and (b), respectively for the case of randomly oriented cubes; panels (c) and (d) show the corresponding data collapses.}
\end{figure}

To achieve an accuracy standard of one part in $10^{3}$ for $\eta_{c}$,
we calculate $\langle f \rangle$ for on the order of nine evenly spaced densities centered about $\eta_{c}$ and
separated in density by $0.25\%$ of the latter. $\eta_{c}$ and $R_{c}$ may be gleaned from the intersection of $\langle f \rangle$
curves. Alternatively, one may take advantage of the data collapse phenomenon near $\eta_{c}$.  For this purpose,
one plots the disorder averaged percolation probability with respect to $x = L^{1/\nu} (\eta - \eta_{c})$.
Data collapses for $\eta_{c}^{\mathrm{GC}}$ and $\eta_{c}^{\mathrm{V}}$ are shown in panel (c) and panel (d)
of Fig.~\ref{fig:Fig2}, respectively for the case of randomly oriented cube-shaped grains.  
The premise of the data collapse is that the Monte Carlo data lies on a
universal curve $g(x)$, and one may use this phenomenon as a quantitative tool to find $\eta_{c}$ and $\nu$.

With an approach similar to that described elsewhere~\cite{djpriour}, we use a high order polynomial to
represent the universal scaling curve with $g(x) = \sum_{j = 0}^{n} A_{j} x^{j}$ with the $A_{j}$ coefficients
fixed by linear least squares fitting; $\eta_{c}$ and $\nu$ are tuned to minimize the chi square
deviation measure.  To take into consideration corrections to scaling and access the bulk  
limit, we fit $\eta_{c}(L)$ and $R_{c}(L)$ to $\eta_{c}(L) = \eta_{c} + A L^{-\delta_{\eta}}$ and 
$R_{c}(L) = R_{c} + B L^{-\delta_{R}}$; $\delta_{\eta}$ and $\delta_{R}$ are 
exponents, and the coefficients $A$ and $B$ are minimized by an optimal choice of $\chi$, 
typically on the order of $\chi = -1.5\rho^{-1/3}$.  In practice, the variation of $\eta_{c}(L)$ and 
$R_{c}(L)$ with respect to system size is modest and is generally within Monte Carlo statistical errors or nearly so.  
Moreover, we invariably find critical indices obtained in this manner to be in agreement with the 
results of global data collapses using Monte Carlo data for all system sizes considered.

Table~\ref{tab:Tab1} contains $\eta_{c}$, $R_{c}$, and $\nu$ for grain cluster and void percolation 
transitions.  On the other hand, Table~\ref{tab:Tab2} shows results for the former in the case of the complementary 
technique of seeking the percolation transition of intersecting or contiguous faces with 
as a marker for grain-cluster percolation with all results in agreement with the 
corresponding $\eta_{c}^{\mathrm{GC}}$ values recorded in Table~\ref{tab:Tab1}.  
Apart from the case of randomly oriented tetrahedra, our $\eta_{c}^{\mathrm{GC}}$ critical grain 
concentration results are in good general agreement with values reported in the literature
for randomly oriented Platonic solids~\cite{Stanley,Torquato,Torquato2,Hyytia,koza2}
as well as aligned cubes.

In the case of grain-cluster percolation, critical densities for aligned and randomly
oriented grains are clearly seen to be distinct, including for the quasi-spherical 
truncated icosahedra.  A salient pattern is the greater $\eta_{c}^{\mathrm{GC}}$ for 
aligned versus randomly oriented inclusions in all cases except for tetrahedra.  We 
surmise that randomly oriented figures may be more likely to disrupt each other's 
volumes than in the aligned case, with smaller critical densities thus needed for the 
presence of system spanning inclusion clusters.  However, this intuition does not account for 
the inverted order in the case of tetrahedra.  A possible explanation is that in the 
aligned case apexes of tetrahedra are optimally positioned to pass through the triangular bases of 
tetrahedra above them.  

In the case of void percolation, there is likewise a consistent pattern in that $\eta_{c}^{\mathrm{V}}$ is 
greater for aligned than randomly oriented grains for polyhedra in which facets capping the figure both 
above and below (i.e. relative to the axis of symmetry) are parallel as is true for 
cubes, dodecahedra, and truncated icosahedra.  It is possible that parallel facets at the top and base of 
neighboring polyhedra may sandwich and leave intact narrow corridors of empty space that might 
otherwise be disrupted.

For octahedra and icosahedra, on the other hand, $\eta_{c}^{\mathrm{V}}$ for randomly oriented figures 
exceeds the critical density for aligned grains.  The concentration thresholds in the case of tetrahedra where only the 
base of each shape is perpendicular to the axis of symmetry in the case of aligned polyhedra are close 
to one another with $\eta_{c}^{\mathrm{V}}$ only
slightly greater for aligned than for randomly oriented grains.

\begingroup
\squeezetable
\begin{table}[h]
\centering
\begin{tabular}{| c | c | c | c | c | c | c |}
\hline
\hline
$\textrm{Grain}$ & $\eta_{c}^{\mathrm{GC}}$ & $\nu_{\mathrm{GC}}$ & $R_{c}^{\mathrm{GC}}$ & $\eta_{c}^{\mathrm{V}}$ & $\nu_{\mathrm{V}}$ & $R_{c}^{\mathrm{V}}$  \\
\hline
$\textrm{Tet}_{\mathrm{A}}$ & 0.13436(5) & 0.90(4) & 0.83(1) &  2.836(2) & 0.88(5) & 0.83(1) \\
$\textrm{Tet}_{\mathrm{R}}$ & 0.16643(5) & 0.87(5) & 0.83(1) &  2.830(2) & 0.96(9) & 0.83(1)  \\
$\textrm{Cub}_{\mathrm{A}}$ & 0.3247(2) & 0.89(2) & 0.83(1) &  3.279(2) & 0.90(5) & 0.82(1) \\
$\textrm{Cub}_{\mathrm{R}}$ & 0.2445(2) & 0.93(3) & 0.84(1) &  3.113(2) & 0.87(5) & 0.83(1) \\
$\textrm{Oct}_{\mathrm{A}}$ & 0.3272(1) & 0.88(4) & 0.83(1) &  3.207(2) & 0.77(13) & 0.82(1) \\
$\textrm{Oct}_{\mathrm{R}}$ & 0.2517(3) & 0.88(3) & 0.83(2) &  3.252(3) & 0.99(18) & 0.83(1)\\
$\textrm{Dod}_{\mathrm{A}}$ & 0.3385(1) & 0.88(3) & 0.83(1) &  3.348(2) & 0.85(7) & 0.82(1) \\
$\textrm{Dod}_{\mathrm{R}}$ & 0.2987(1) & 0.90(2) & 0.82(1) &  3.339(2) & 0.83(8) & 0.82(1) \\
$\textrm{Ico}_{\mathrm{A}}$ & 0.3393(5) & 0.89(5) & 0.83(1) &  3.381(3) & 0.87(7) & 0.82(1) \\
$\textrm{Ico}_{\mathrm{R}}$ & 0.3054(3) & 0.85(5) & 0.82(1)) & 3.414(3) & 0.91(5) & 0.83(1) \\
$\textrm{Tr Ico}_{\mathrm{A}}$ & 0.3414(2) & 0.88(4) & 0.83(1) & 3.459(2) & 0.91(5)  & 0.82(1) \\
$\textrm{Tr Ico}_{\mathrm{R}}$ & 0.3263(2) & 0.88(4) & 0.83(1)&  3.452(2) & 0.91(6) & 0.82(1) \\
\hline
\hline
\end{tabular}
\caption{\label{tab:Tab1} Critical indices calculated by identifying void volumes for $\eta_{c}^{\mathrm{GC}}$
and $\eta_{c}^{\mathrm{V}}$.  Abbreviations (e.g. ``Cub'' for cubes) indicate
the grain shapes, while subscripts ``A'' and ``R'' indicate aligned and randomly oriented cases
respectively, and ``Tr'' indicates a truncated figure.}
\end{table}
\endgroup

\begingroup
\begin{table}[h]
\centering
\begin{tabular}{|c|c|c|c|}
\hline
\hline
$\textrm{Grain}$ & $\eta_{c}^{\mathrm{VC}}$ & $\nu_{\mathrm{GC}}$ & $R_{c}^{\mathrm{GC}}$ \\
\hline
$\textrm{Tet}_{\mathrm{A}}$ & 0.1344(1) & 0.86(2) & 0.82(1) \\
$\textrm{Tet}_{\mathrm{R}}$ & 0.16640(15) & 0.86(3) & 0.81(1) \\
$\textrm{Cub}_{\mathrm{A}}$ & 0.3247(2) & 0.88(2) & 0.82(1) \\
$\textrm{Cub}_{\mathrm{R}}$ & 0.2443(3) & 0.88(1) & 0.82(1) \\
$\textrm{Oct}_{\mathrm{A}}$ & 0.3273(3) & 0.86(2) & 0.84(2) \\
$\textrm{Oct}_{\mathrm{R}}$ & 0.2517(3) & 0.89(2) & 0.84(1) \\
$\textrm{Dod}_{\mathrm{A}}$ & 0.3384(6) & 0.85(2) & 0.81(2) \\
$\textrm{Dod}_{\mathrm{R}}$ & 0.2986(3) & 0.87(2) & 0.82(1) \\
$\textrm{Ico}_{\mathrm{A}}$ & 0.3396(3) & 0.89(2) & 0.84(1) \\
$\textrm{Ico}_{\mathrm{R}}$ & 0.3056(3) & 0.88(2) & 0.83(1) \\
$\textrm{Tr Ico}_{\mathrm{A}}$ & 0.3413(3) & 0.87(1) & 0.83(1) \\
$\textrm{Tr Ico}_{\mathrm{R}}$ & 0.3263(4) & 0.88(2) & 0.82(1) \\
\hline
\hline
\end{tabular}
\caption{\label{tab:Tab2} Critical indices calculated using neighboring planes for $\eta_{c}^{\mathrm{GC}}$.
Abbreviations (e.g. ``Cub'' for cubes) indicate the grain shapes, while subscripts ``A`` and ``R'' indicate 
aligned and randomly oriented cases, respectively, and ``Tr'' indicates a truncated figure.}
\end{table}
\endgroup

\section{Structurally Disordered Grains}

Finally, we calculate $\eta_{c}^{\mathrm{GC}}$ and $\eta_{c}^{\mathrm{V}}$ in the context of structural disorder 
in the form of aggressively fragmented constituent grains.  
The latter are subject to a sequence of randomly oriented and randomly placed slicing planes which successively 
cleave away material. The slice removed is the portion of the figure 
which does not contain the origin at the center of the original cube.  
The number of fractures imposed per grain is sampled from Poissonian statistics based on a tunable 
number of slices per unit volume.  However, as a more pertinent parameter for the structural disorder strength, we 
use $N_{\mathrm{sust}}$ (calculated a posteriori by sampling $10^{9}$ grains generated 
by the slicing process), the mean number of slices that are sustained in the sense of cleaving away at least one vertex.
The mean inclusion volume $\langle v_{\mathrm{B}} \rangle$ used, e.g., in $\eta_{c} = \rho_{c} \langle v_{\mathrm{B}} \rangle$ 
is calculated by summing the volumes of each of the component tetrahedra defined by the face center, and the two 
vertices of a facet edge; as in the case of $N_{\mathrm{sust}}$, we average over a billion realizations of 
disorder.  

\begin{figure}
\includegraphics[width=.5\textwidth]{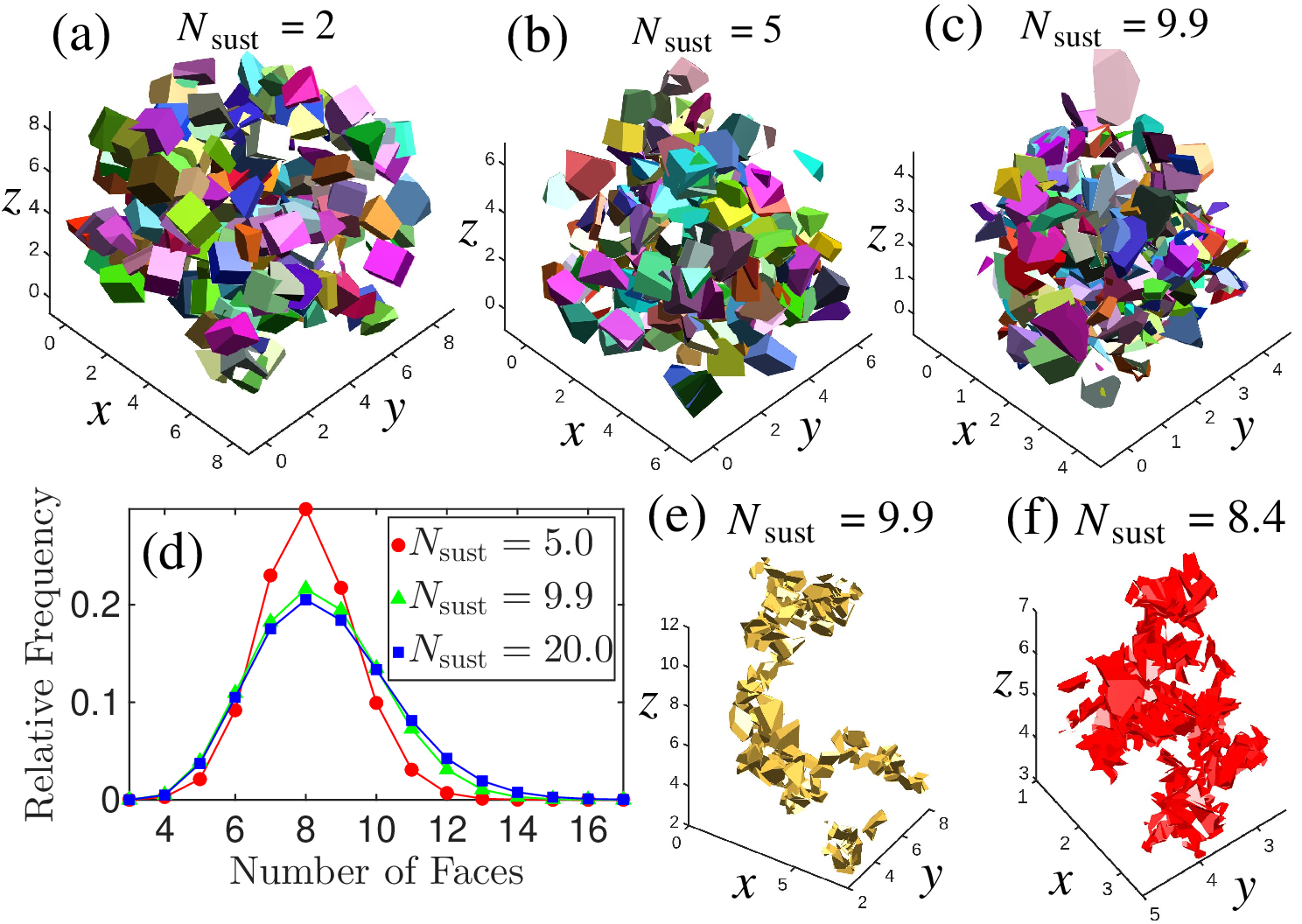}
\caption{\label{fig:Fig3} (Color online) Panel (a), (b), and (c) depict sample assemblies of
grains for various numbers of sustained slices $N_{\mathrm{sust}}$.  The graph in panel (d) is a 
frequency plot for facet number for a range of $N_{\mathrm{sust}}$ values.  Panels (e) and (f) show
a portion of a percolating free surface near $\eta_{c}^{\mathrm{GC}}$ and $\eta_{c}^{\mathrm{V}}$,
respectfully, for the $N_{\mathrm{sust}}$
values indicated.}
\end{figure}

Panels (a), (b), and (c) of Fig.~\ref{fig:Fig3} show assemblies of structurally disordered grains for
$N_{\mathrm{sust}} = 2$, $N_{\mathrm{sust}} = 5$, and $N_{\mathrm{sust}} = 9.9$, respectively.  
As may be seen in the frequency plot in panel (d) of Fig.~\ref{fig:Fig3} after many fracture events, the frequency distribution of the number of 
facets per grain tends to a limiting profile as $N_{\mathrm{sust}}$ exceeds on the order of 10 
with a peak at eight faces.  This typically modest number of faces even after many 
sustained slices makes finding void volumes tractable even for very agressively fragmented inclusions.  To capitalize 
on the small number of facets for typical fragments, an inventory of planar faces, edges, and vertices is maintained and dynamically 
updated as each slice is imposed.  Ultimately, only vertices which have not been sheared away remain to be considered, 
significantly facilitating the 
task of identifying the void volumes and allowing for examination of $N_{\mathrm{sust}} \gg 1$ where the $\eta_{c}^{\mathrm{GC}}$ and 
$\eta_{c}^{\mathrm{V}}$ saturate with respect to the mean number of fracturing events per inclusion.
In a similar manner to Fig.~\ref{fig:Fig1}, panels (e) and (f) of Fig.~\ref{fig:Fig3} show worm-like inclusions clusters
near $\eta_{c}^{\mathrm{GC}}$ and tunnel like void volumes near $\eta_{c}^{\mathrm{V}}$, respectively, for 
aggressively fractured inclusions. 

\begin{figure}
\includegraphics[width=.45\textwidth]{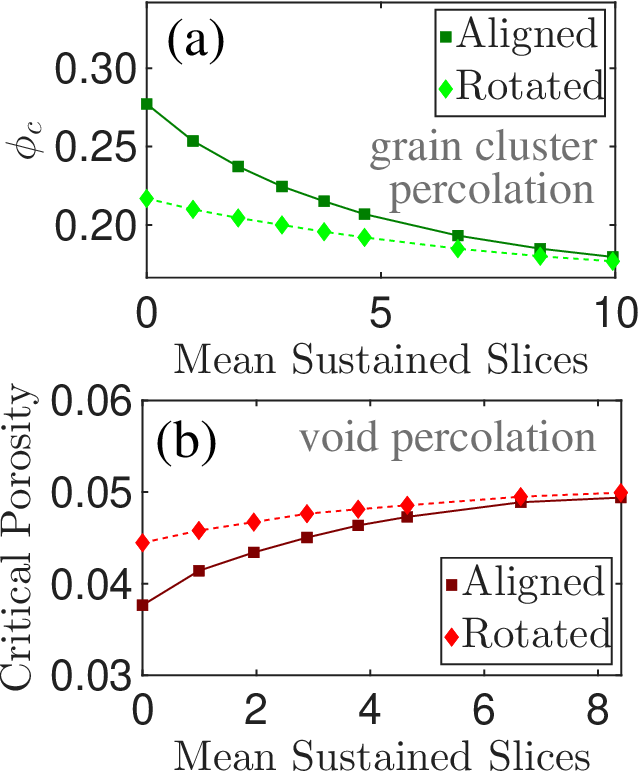}
\caption{\label{fig:Fig4} (color online) Critical concentrations are shown in panels
(a) and (b) in the case of grain cluster 
and void percolation, respectively for the case of structurally disordered grains.  
The vertical axis for the former is the critical excluded volume 
$\phi_{c} = 1 - e^{-\eta_{c}^{\mathrm{GC}}}$ and the critical porosity $e^{-\eta_{c}^{\mathrm{V}}}$
for the latter.
Monte Carlo statistical errors are smaller than the symbol sizes.}
\end{figure}

To facilitate extrapolation to the $N_{\mathrm{sust}} \gg 1$ limit, we performed independent calculations for aligned and randomly 
oriented cubes which are then subject to a sequence of random fractures.  Coupled with this choice is the expectation that the 
disappearance of the gap among $\eta_{c}$ for aligned versus randomly oriented shapes, indicating a loss of the ``memory'' 
of the original solids,  will coincide with a saturation of both $\eta_{c}$ curves at a limiting critical concentration.

Fig.~\ref{fig:Fig4} shows results for both grain cluster percolation and void percolation in the case of structurally disordered 
fragments in panel (a) and panel (b), respectively.
In the case of the former, the vertical axis is the excluded volume $\phi_{c} = 1 - e^{-\eta_{c}^{\mathrm{GC}}}$,
while for the latter the critical porosity (i.e. $e^{-\eta_{c}^{\mathrm{V}}}$) is shown.  While $\phi_{c}$ for
grain cluster percolation results appears not to have fully saturated even for 10 sustained slices per grain on average,
the critical porosity for void percolation does appear to level out at 5\% near $N_{\mathrm{sust}} = 8$. 

\section{Conclusions}

In conclusion, we have developed a geometrically exact technique for identifying void volumes a priori 
and determining if interstitial volume networks percolate. With large-scale Monte Carlo calculations, we have 
significantly improved the accuracy of $\eta_{c}^{\mathrm{V}}$ results in the case of the Platonic solids,
to the degree that differences among void percolation thresholds for aligned versus randomly oriented 
grains is resolved.  

For the sake of taking into consideration the structurally disordered and aggressively fragmented grains one finds 
in nature, we have examined cube-shaped inclusions subject to randomly oriented slicing planes, finding a 
convergence to 5\% critical porosity in the limit of many accumulated fractures per grain.


\begin{acknowledgments}
We acknowledge helpful discussions with Michael Crescimanno.
\end{acknowledgments}



\begin{thebibliography}{11}
\expandafter\ifx\csname natexlab\endcsname\relax\def\natexlab#1{#1}\fi
\expandafter\ifx\csname bibnamefont\endcsname\relax
  \def\bibnamefont#1{#1}\fi
\expandafter\ifx\csname bibfnamefont\endcsname\relax
  \def\bibfnamefont#1{#1}\fi
\expandafter\ifx\csname citenamefont\endcsname\relax
  \def\citenamefont#1{#1}\fi
\expandafter\ifx\csname url\endcsname\relax
  \def\url#1{\texttt{#1}}\fi
\expandafter\ifx\csname urlprefix\endcsname\relax\def\urlprefix{URL }\fi
\providecommand{\bibinfo}[2]{#2}
\providecommand{\eprint}[2][]{\url{#2}}

\bibitem{stauffer} D. Stauffer and A. Aharony, \textit{Introduction to Percolation Theory},
2nd ed. (Taylor and Francis, Bristol, 1994).

\bibitem{martys} N.~S.~Martys, S~Torquato, and D.~P.~Bentz, 
Universal scaling of fluid permeability for sphere packings,
Phys. Rev. E \textbf{50}, 403 (1994).

\bibitem{maier} R.~S.~Maier, D.~M.~Kroll, H.~T.~Davis, and R.~S.~Bernard, Diffusion and Flow in 
Porous Domains Constructed Using Process-Based and Stochastic Techniques, J. Colloid Interface Sci.
\textbf{217}, 341 (1999).

\bibitem{Yi1} Y.~B.~Yi, Void percolation and conduction of overlapping ellipsoids,
Phys. Rev. E \textbf{74}, 031112 (2006).

\bibitem{Yi2} Y.~B.~Yi and K.~Esmail, Computational measurement of void percolation 
thresholds of oblate particles and thin plate composites, J. Appl. Phys. \textbf{111}, 124903 (2012).

\bibitem{koza} Z.~Koza, G.~Kondrat, and K.~Suszcy\'{n}ski, Percolation of overlapping 
squares or cubes on a lattice, J. Stat. Mech. (2014) P11005.

\bibitem{elam} W.~T.~Elam, A.~R.~Kerstein, and J.~J.~Rehr, 
Critical properties of the void percolation problem for spheres,
Phys. Rev. Lett.
\textbf{52}, 1516 (1984).

\bibitem{marck} S.~C.~van der Marck, Network Approach to Void Percolation in 
a Pack of Unequal Spheres, Phys. Rev. Lett. \textbf{77}, 1785 (1996).

\bibitem{rintoul} M.~D.~Rintoul, Precise determination of the void percolation 
threshold for two distributions of overlapping spheres, Phys. Rev. E \textbf{62}, 68 (2000).

\bibitem{klatt} M.~A.~Klatt, R.~M.~Ziff, and S.~Torquato, 
Critical pore radius and transport properties of disordered hard- and 
overlapping-sphere models,
Phys. Rev. E \textbf{104}, 014127 (2021).

\bibitem{beijeren} H.~van Beijeren, 
Transport properties of stochastic Lorentz models,
Rev. Mod. Phys. \textbf{54}, 195 (1982).

\bibitem{bauer} T.~Bauer, F.~H\"{o}fling, T.~Munk, E.~Frey, and T.~Franosch,
The localization transition of the two-dimensional Lorentz model,
Eur. Phys. J. Spec. Top. \textbf{189}, 103 (2010).

\bibitem{hofling} F.~H\"{o}fling, T.~Munk, E.~Frey, and T.~Franosch, 
Critical dynamics of ballistic and Brownian particles in a heterogeneous environment,
J. Chem. Phys. 
\textbf{128}, 164517 (2008).

\bibitem{spanner} M.~Spanner, F.~H\"{o}fling, S.~C.~Kapfer, K.~R.~Mecke, 
G.~E.~Schr\"{o}der-Turk, and T.~Franosch, 
Splitting of the Universality Class of Anomalous Transport in Crowded Media,
Phys. Rev. Lett. \textbf{116}, 060601 (2016).

\bibitem{hofling2} F.~H\"{o}fling, T.~Franosch, and E.~Frey, 
Localization Transition of the Three-Dimensional Lorentz Model and Continuum Percolation,
Phys. Rev. E \textbf{96},
165901 (2006).

\bibitem{kammerer} A.~Kammerer, F.~H\"{o}fling, and T. Franosch, 
Cluster-resolved dynamic scaling theory and universal corrections for transport on 
percolating systems,
Europhys. Lett. 
\textbf{84}, 66002 (2008).

\bibitem{spanner2} M.~Spanner, F.~H\"{o}fling, G.~E.~Schr\"{o}der-Turk, K.~Mecke, 
and T.~Franosch, 
Anomalous transport of a tracer on percolating clusters,
J. Phys. Condens. Matter \textbf{23}, 234120 (2011).

\bibitem{djpriour} D.~J.~Priour, Jr., 
Percolation through voids around overlapping spheres:
A dynamically based finite-size scaling analysis,
Phys. Rev. E \textbf{89}, 012148 (2014).

\bibitem{djpriour2} D.~J.~Priour and N.~J.~McGuigan, 
Percolation through Voids around Randomly Oriented Polyhedra and Axially Symmetric Grains,
Phys. Rev. Lett. \textbf{121}, 225701 (2018).

\bibitem{ballow} A.~Ballow, P.~Linton, and D.~J.~Priour, Jr., 
Percolation through voids around toroidal inclusions,
Phys. Rev. E \textbf{107}, 014902 (2023).

\bibitem{Skolnick} M.~Skolnick and S.~Torquato, 
Accurate formula for the effective conductivity of highly clustered
two-phase materials,
Physical Review Materials \textbf{9}, 055601 (2025).

\bibitem{polyvolume} E.~W.~Weinstein, \textit{CRC Concise Encyclopedia of Mathematics}, 
2nd ed. (Chapman \& Hall/CRC, Boca Raton, FL, 2003).

\bibitem{bruin} C.~Bruin, 
A computer experiment on diffusion in the Lorentz gas,
Physica (Amsterdam) \textbf{72}, 261 (1974).

\bibitem{hoshen} J.~Hoshen and R.~Kopelman, 
Percolation and cluster distribution. I. Cluster multiple
labeling techniques and critical concentration algorithm,
Phys. Rev. B \textbf{14}, 3438 (1976).

\bibitem{wang} J.~Wang, Z.~Zhou, W.~Zhang, T.~M.~Garoni, and Y.~Deng, 
Bond and site percolation in three dimensions,
Phys. Rev. E
\textbf{87}, 052107 (2013).

\bibitem{Stanley} Don~R.~Baker, Gerald Paul, Sameet Sreenivasan, and H.~Eugene Stanley,
Continuum percolation threshold for interpenetrating squares and cubes,
Phys. Rev. E \textbf{66}, 046136 (2002).

\bibitem{Torquato} S.~Torquato and Y.~Jiao, 
Effect of dimensionality on the percolation threshold of overlapping nonspherical hyperparticles,
Phys. Rev. E \textrm{87}, 022111 (2012). 

\bibitem{Torquato2} S.~Torquato and Y.~Jiao, 
Effect of dimensionality on the continuum percolation of overlapping hyperspheres and hypercubes. 
II.  Simulation results and analyses,
J. Chem. Phys. \textrm{137}, 074106 (2012).

\bibitem{Hyytia} E.~Hyyti\"{a} J.~Virtamo, P.~Lassila, and J.~Ott, 
Continuum Percolation Threshold for Permeable Aligned Cylinders and Opportunistic Networking,
IEEE Communications Letter.
\textrm{16}, 1064 (2012).

\bibitem{koza2} Zbigniew Koza and Jakub Pola, 
From discrete to continuous percolation in dimensions 3 to 7,
Journal of Statistical Mechanics:
Theory and Experiment \textbf{2016}, 103206 (2016).

\end{thebibliography}
\end{document}